# Correlative Raman Imaging and Scanning Electron Microscopy: the Role of Single Ga Islands in Surface - Enhanced Raman Spectroscopy of Graphene

*Jakub Piastek [1,2], Jindřich Mach*[\*,1,2], Stanislav Bardy [2], Zoltán Édes[1,2], Jaroslav Maniš[1] Vojtěch Čalkovský[1], Miroslav Bartošík[1,2,3], Martin Konečný[1,2], Jiří Spousta, and omáš Šikola [1,2]*

(1) CEITEC BUT, Brno University of Technology, Technická 3058/10, 616 00 Brno, Czech Republic

(2) Institute of Physical Engineering, Brno University of Technology, Technická 2, 616 69 Brno, Czech Republic

(3) Department of Physics and Materials Engineering, Faculty of Technology, Tomas Bata University in Zlín, 760 01, Czech Republic

KEYWORDS: Graphene, Ga, deposition, SERS, Raman spectroscopy, RISE, SEM

ABSTRACT:

*Surface enhanced Raman spectroscopy (SERS) is a perspective non-destructive analytic technique enabling detection of individual nanoobjects, even single-molecules. . In the paper, we have studied the morphology of Ga islands deposited on CVD graphene by ultrahigh vacuum (UHV) evaporation and local optical response of this system by the correlative Raman Imaging and Scanning Electron Microscopy (RISE). Contrary to the previous papers, where*

*only an integral Raman response from the whole ununiformed Ga NPs ensembles on graphene was investigated, the RISE technique has enabled us to detect graphene Raman peaks enhanced by single Ga islands and particularly to correlate the Raman signal with the shape and size of these single particles. In this way and by a support of numerical simulations, we have proved a plasmonic nature of the Raman signal enhancement related to localized surface plasmon resonances (LSPR).*

*It has been found that this enhancement is island-size dependent and shows a maximum for medium-sized Ga islands. A reasonable agreement between the simulations of the plasmon enhancement of electric fields in the vicinity of Ga islands and the experimental intensities of corresponding Raman peaks proved the plasmonic origin of the observed effect known as the Surface Enhanced Raman Spectroscopy.*

Enhanced near fields of localized surface plasmons in metal nanostructures form a basis of a wide-spread non-destructive analytic method known as Surface Enhanced Raman Spectroscopy (SERS) enabling high sensitive detection down to single molecular level [1–3].

The Raman scattering itself has a very low cross section ($\approx 10^{-28}$ cm$^2$.sr$^{-1}$) and thus provides limited sensitivity and applicability. Nevertheless, it has been successfully utilized for characterization and analysis of two-dimensional graphene structures by means of three most distinctive characteristic peaks (G, D, 2D). These peaks give us significant information about the quality, thickness, electronic structure and induced stress of graphene. The **G** peak ( ~ 1585 cm$^{-1}$) is associated with a degenerate in-plane mode $E_{2g}$, at the Γ point in the Brillouin zone, and is the only band coming from the normal first order Raman scattering process in graphene [4]. Therefore, its intensity is a characteristic feature and can be used for the specification of the strain induced by an external force on graphene [5]. The **D** peak (~ 1345 cm$^{-1}$) appears when defects in graphene are present. The 2D peak (~ 2670 cm$^{-1}$) comes from a double resonance. This mechanism includes two phonons near the K point that is strongly dependent on the electronic and atomic structure of graphene[6]. The Raman signal enhancement from graphene can be realized by the choice of a suitable substrate - for example, a commonly utilized substrate $SiO_2$ (280 nm)/Si enhances the signal up to 30 times due to an interference-enhanced Raman scattering (IERS) effect [7]. As it has been shown in the former studies, SERS on a graphene sheath can be also realized by employing metal nanostructures. In this case, metals as Au, Ag and Cu are very often used to achieve a significant enhancement. However, this enhancement is mostly accompanied by a high noise, particularly caused by fluorescence which is generally much bigger than the Raman scattering (the fluorescence cross section reaches approx. $10^{-16}$ $cm^2.sr^{-1}$). Moreover, using these metals a significant instability is observed due to the catalytic effects ongoing on metal surfaces [8–11]. In contrast, it has been found that graphene is capable of quenching fluorescence of molecules, thus facilitating detection of the Raman signal. It is possible to prepare hybrid systems acting as efficient SERS substrates combining graphene unique properties with plasmonic active metals [12]. Contrary to Au and Ag, Ga is not a catalytically active metal and, in addition, it is covered by a stable oxide layer [13]. Similarly to a $SiO_2$ shell of some metal nanoparticles [6], such a layer might result in a more profound SERS signal. The SERS effect at a Ga/graphene interface has been already reported in [14]. In that paper, the SERS platform was formed by graphene and Ga nanoparticles (NPs) deposited at RT. It has been shown that the enhancement factor (defined as the Raman intensity for Ga NPs on graphene divided by Raman intensity for pristine graphene) for the G- and 2D peaks grows with the increasing nanoparticle size up to the value 200. However, in this work only an integral Raman response from the whole Ga NPs ensembles on graphene was investigated. The distribution of the size of Ga nanoparticles was not uniformed and this size was characterized by mean values of NPs radius. The graphene – Ga samples possessing only three different mean values of NPs radius (8 nm, 35 nm, and 50 nm) were investigated. Hence, the direct relationship between the Raman intensity and the size of individual nanoparticles was not available. Thus the explanation of the Raman signal enhancement by the effect of localized surface plasmon resonances in Ga NPs was not supported by a more detailed analysis and was rather speculative.

In our work, we directly tackle this challenge by the application of correlative Raman Imaging and Scanning Electron microscopy (RISE) to Ga islands grown on a graphene/SiO2 substrate at different substrate temperatures (RT-500 ºC). By this method, we have correlated the Raman signal collected from the areas of single Ga islands with the shape and size of these particles. In this way and by a support of numerical simulations, we have proved a plasmonic nature of the Raman signal enhancement related to localized surface plasmon resonances (LSPR).

It is worth mentioning that gallium on graphene is a proper testing system for investigation of SERS, since Ga atoms weakly bind to graphene and do not perturb to a great extent its electronic structure [15],[16]. Thus, there is no strong hybridization between Pz orbitals of graphene and valence electrons of Ga [17] which should contribute to better stability and interpretation of the Raman signal enhancement.

**RESULTS AND DISCUSSION**

*Morphology of Ga layers on graphene*

Deposition of Ga atoms on CVD single domain graphene flakes was carried out at different substrate temperatures. The goal was to prepare samples with Ga islands - droplets on graphene possessing a sufficient mutual separation and enabling thus to take the Raman signal from single Ga islands. SEM images of the samples after Ga deposition are shown in Fig.1. To provide direct comparison, there are simultaneously shown two different types of surfaces on each sample - a single domain graphene flake and a bare $SiO_2$ area. The highest concentartion of Ga islands is observed on the sample prepared by Ga deposition at the substrate temperature $T = 25^{\circ}C$ both for graphene and $SiO_2$ surfaces, see Fig.1a. The Ga islands form droplets of a spherical shape, most likely truncated in their lower part by the substrate (for simplicity we will call them sphere-like islands). The Ga island size and their separation grow with an increasing substrate temperature due to higher thermal energy of Ga atoms and thus more intense Ostwald ripening resulting in formation of bigger Ga islands at the expense of smaller ones. Such a behavior can be observed for the samples prepared at the substrate temperature $T = 200^{\circ}C$ in Fig.1b. A further increase of temperature above $200^{\circ}C$ leads to the formation of two types of Ga islands – in addition to gallium sphere-like islands, also gallium flat islands appear, as can be seen in Fig. 1c, where a SEM image of the sample surface prepared at a temperature of $300^{\circ}C$ is shown. The flat islands are lower and usually much wider (several hundreds of nanometers) than the Ga spheres-like islands (see

Fig. S1 in Supplementary Information). In the SEM images they appear gray because they are thinner and therefore provide a lower yield of secondary electrons. The results of energy-dispersive X-ray spectroscopy (EDX) measurements proved that both the sphere-like and flat-like islands consist of gallium.

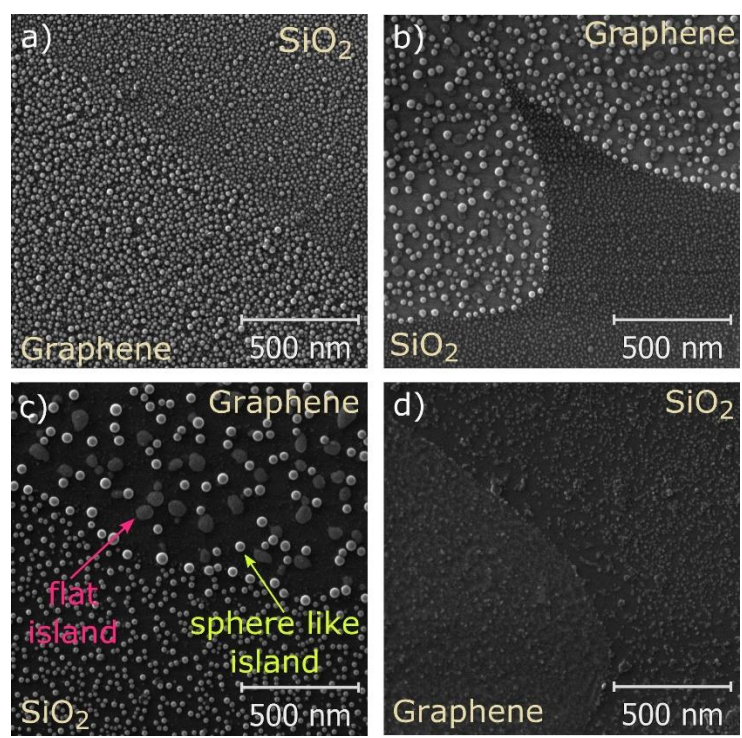

**Fig 1.** SEM images of Ga islands grown on graphene single domain flakes transferred on the SiO$_2$ layer at a) 25°C, b) 200°C, c) 300°C, and d) 500°C. Deposition rate - 0.2 ML/min, deposition time - 60 minutes.

The mechanism of the Ga flat island growth is not exactly known to us yet. One can deduce these two distinct types of structures could appear due to local inhomogeneities of graphene – lattice distortions and impurities which probably arise after the graphene transfer on the SiO$_2$ layer. They might locally change the wetting properties of the surface which could lead to the change of formation of Ga islands.

At 500°C and higher temperatures, gallium islands disappeared as all the impinging gallium atoms desorbed from the surface (see Fig.1 d).

*2.2 Correlative microscopy of Ga islands - Raman Imaging and SEM (RISE)*

Gallium islands grown at 350°C substrate temperature were visible even by optical microscopy (Fig.2a). In accordance with our findings described above, it was possible to distinguish two types of structures – sphere-like and flat islands, visible by SEM as the bright spots in b) and c). The flat islands having mostly larger diameters and approximately half of the spheres heights. The profiles of typical spheres-like and flat island measured by AFM are in Fig.2 d).

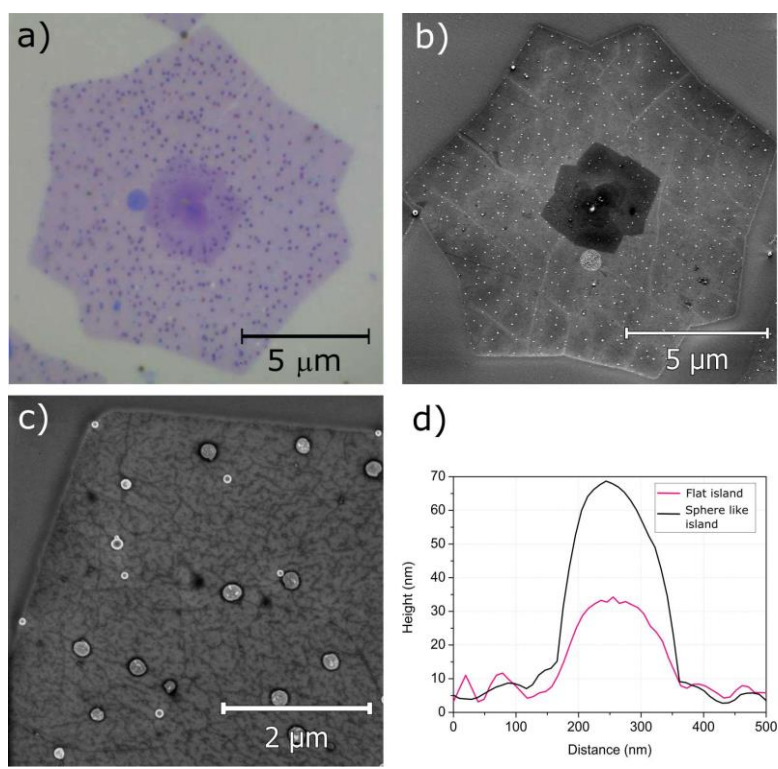

**Fig. 2**. Images of large Ga islands deposited on a graphene single domain flake for 60 min at 350°C: a) bright field optical image, b)-c) SEM images, d) AFM-height profiles of a sphere and disk of Ga.

The sample with large Ga islands grown at 350°C was investigated by the correlative Raman imaging and SEM (RISE) technique. The surface distribution – map of the G peak (1585 cm$^{-1}$) intensities is shown in Fig. 3a. It covers an area of 5 x 5 μm$^2$, and consists of 50 by 50 pixels. SEM scans of the mapped area were taken before and after the μRaman measurements to be sure Ga islands were left intact. As can be seen from the correlation of the Raman map and SEM image, the investigated area contains $SiO_2$ in the top left corner and the rest is part of the graphene flake covered with Ga islands (Fig. 3c). Further, considering the typical extracted spectra shown in Fig. 3b, one can claim there are sites on the sample where all the D, G and 2D peaks of the Raman spectrum were enhanced. By correlating the Raman map with the SEM image acquired at the same place, we were able to identify these sites as Ga islands (Fig.3 c). Almost all the Ga islands (both spheres and disks) but two (encircled in correlated Fig. 3c) contribute to the enhancement of the Raman signal. Interestingly, disks enhance this signal much stronger than the spheres.

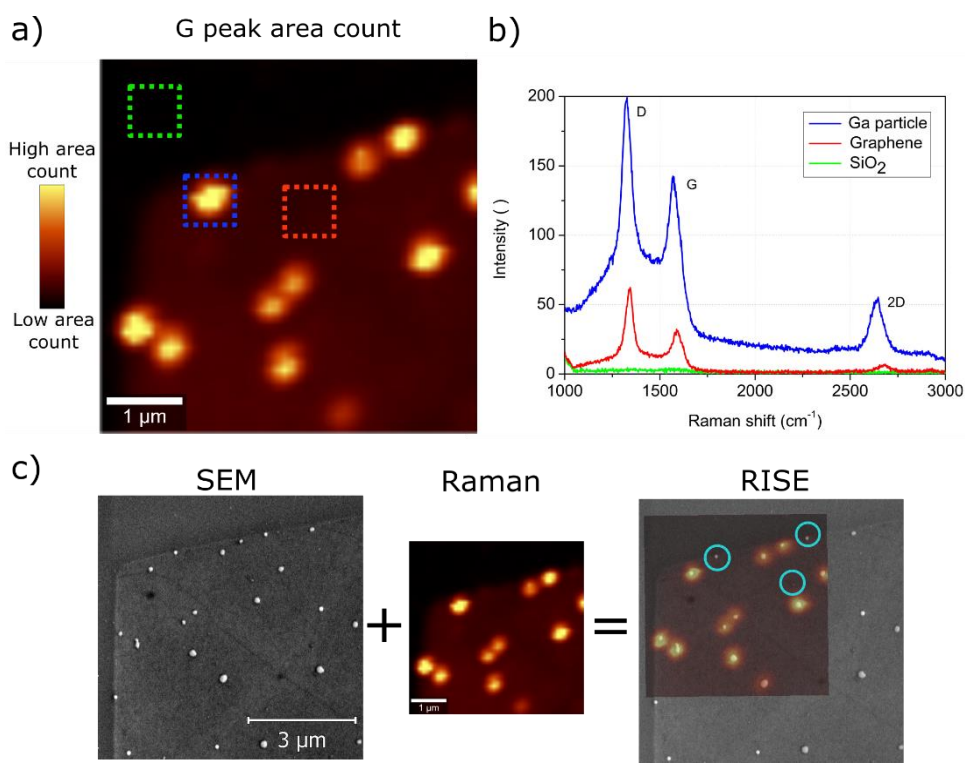

**Fig.3**. a) Raman map of the G peak (1585 cm$^{-1}$) intensity for a sample with large Ga islands on graphene. b) μRaman spectra extracted from the corresponding sites of the Raman map marked by colored squares in a). c) RISE allows us to correlate the SEM image with the Raman map and thus to identify which surface features contribute to the strong Raman signal.

The correlated RISE image in Fig.4 shows how the enhancement of the Raman G peak (1585 cm$^{-1}$) intensity varies from island to island and depends on the island size. The Raman map covers a sample area of 7.5 x7.5 μm$^2$ with the same pixel density as in Fig. 3a) and more than fifty Ga islands in it.

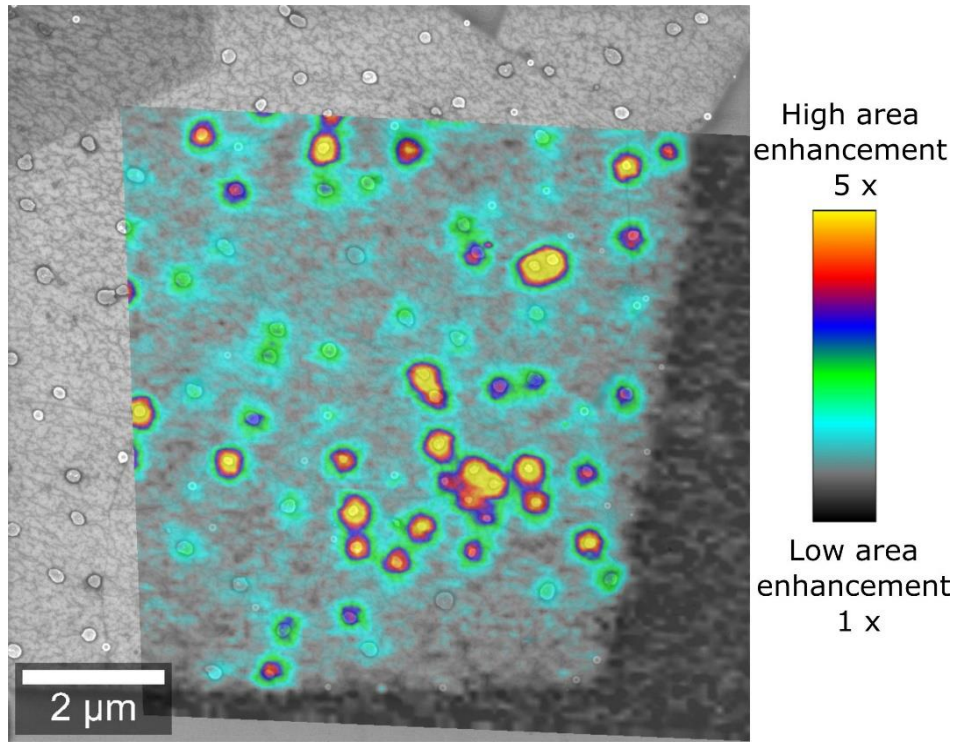

**Fig.4.** Raman map (1585 cm$^{-1}$ G peak enhancement) correlated with SEM images. The pixel number and integration time per spectrum is 75 x 75 px$^2$, and 0.5 s, respectively.

The Raman signal enhancement has been defined as the ratio of the gallium peak intensity measured at a specific point of Ga island- single-layer graphene (SLG) structure to the minimum intensity obtained from this structure between Ga islands. As the peak intensity measure, the height of graphene peaks has been considered. A clear, almost five-fold enhancement in the vicinity of Ga islands has been observed. Compared to the enhancement reported in [18] (200 x), this value seems to be quite small. However, in our case the Raman signal enhancement given just by a single island was measured and the laser beam diameter was at least 300 nm large. It means that the laser was illuminating a much bigger area then that one of a single island, and so the most of the signal was collected just from graphene surface. On top of that, in [14] a dense coverage of graphene surface by Ga particles was used. Hence, all together, these two values are not directly comparable.

One can hypothesize the enhancement of the Raman signal by Ga islands is caused by plasmon effects. The incident laser beam generates local surface plasmons (LSP) in Ga islands which leads upon resonance to a big enhancement of electromagnetic near-field in their vicinity [18]. As a result, the intensity of the Raman signal from the sample area with such a spatially localized intense electric field, called "hot spot" is increased, the effect known as Surface Raman Spectroscopy (SERS).

To prove this hypothesis and shed more light on our results, we carried out numerical simulations of electric field enhancement by Ga islands using a finite-difference-time-domain (FDTD) software (Lumerical). An electromagnetic pulse (plain-waves package) was directed at a model sample with a Ga hemisphere on its top, as depicted in Fig.5. The model sample consisted of a semi-infinite Si substrate consequently covered with a 280 nm $SiO_2$ thin film, SLG of an effective thickness of 0.335 nm, and a Ga hemisphere of different radii ranging from 30 to 120 nm. A dielectric function of Ga was taken from the experimental work published in [19]. The simulated electric field enhancement, related to the electric field value of the incident electromagnetic wave of particular wavelength, is shown as a function of the wavelength of the incident light for different Ga hemisphere radii in Fig.5. The value of the enhanced electric field was read at the point nearby the hemisphere marked by the red cross and called the detection spot in the figure.

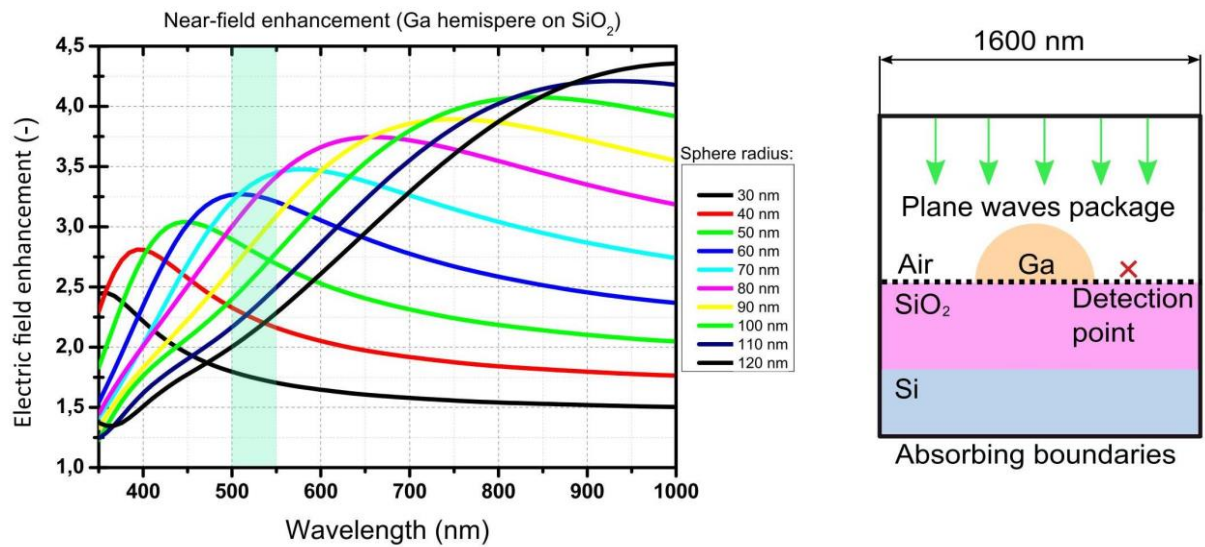

**Fig.5.** Simulated electromagnetic electric field enhancement in the vicinity of a Ga hemisphere of different radii as a function of the incident light wavelength. The green area in the graph is a spectral window around a wavelength of 532 nm (corresponding to the laser applied in μRaman spectroscopy) and used for evaluation of the electric field enhancement. On the right, schematic of the simulation model. The detection point is 10 nm to the right and 5 nm upwards from the Ga/graphene contact point. Simulated by the FDTD Lumerical software.

It is generally assumed that the Raman scattering enhancement scales with the fourth power of the electric field enhancement factor. This assumption is roughly valid if the frequency shifts of Raman peaks from the excitation laser frequency are smaller than the half-width of plasmon resonant peaks of metallic structures [19]. As in our case the maximum shift is less than 90 nm (excitation laser – 532 nm, 2D peak - 620 nm) and the half-width of the resonant peaks is 150 nm and more (see Fig. 5), this condition is valid and we can accept this assumption.

The fourth power of the electric field enhancement corresponding to a narrow spectral range around the Raman excitation laser wavelength (532 nm) is plotted in Fig. 6 together with experimentally measured intensities of D, G and 2D peaks for different Ga sphere radii. Here, the sphere radius represents both the sphere radius of Ga islands determined by SEM and also the radii of Ga hemispheres used in the model. The electric field enhancement was normalized with respect to its maximum, the Raman peak intensities to an average maximum of the Raman peaks calculated in the island radius interval 60 nm – 80 nm. Despite the fact that the real sphere-like Ga islands are covered with a $Ga_2O_3$ layer and approximately modelled with ideal Ga hemispheres, the coincidence between the profiles of Raman peak intensities and electric field enhancement is evident.

To be sure that the signal enhancement is not just given by the effect of Fabry-Perot resonances (multiple-reflection interference) in our multilayer system, we have also carried out numerical simulations (Lumerical) of the signal reflected from a Ga/SiO2/Si layered model (graphene layer was neglected) as the function of the Ga layer thickness. As can be seen in Fig. S2, no clear dependence on the Ga layer thickness has been observed for any wavelength which is quite different from the Raman signal dependence on the Ga island radius shown in Fig. 6.

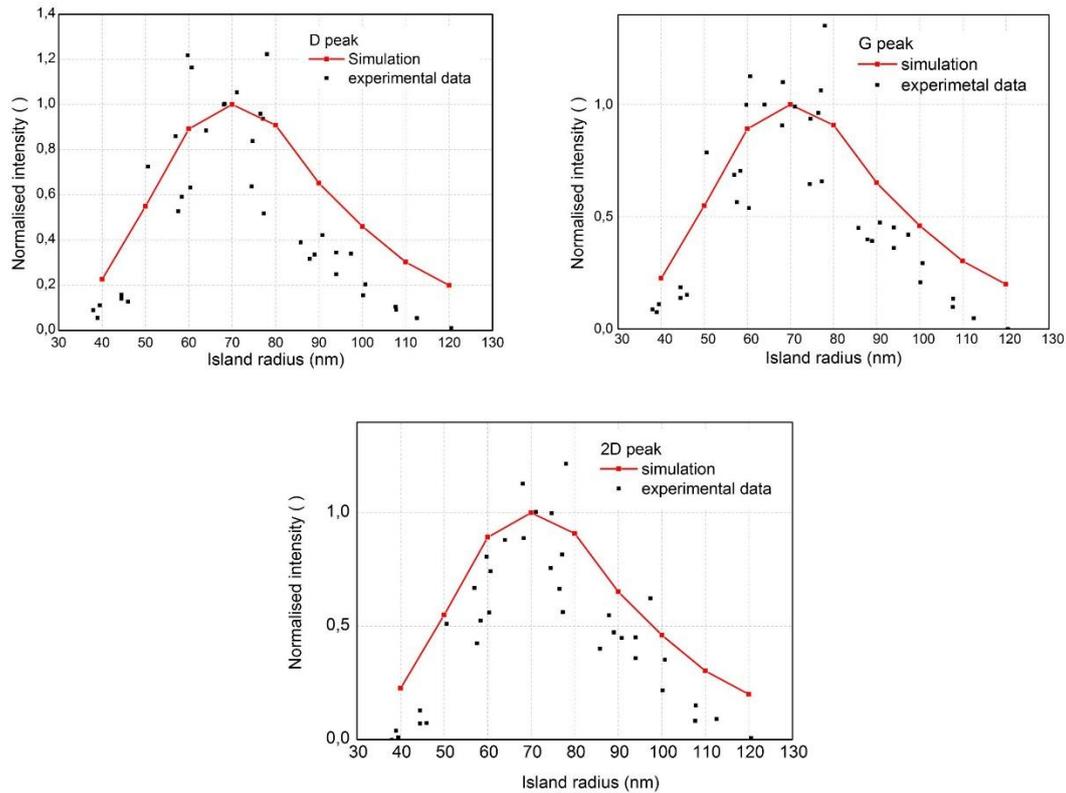

**Fig.6.** Comparison of the measured Raman peak (D, G, 2D) intensities with simulations of the electric field enhancement for different Ga island radii. The red curve connecting the simulated data is drawn for guiding the eye only.

Hence, these results confirm that the Raman peak enhancement is attributed to plasmon electric field enhancement across and in the vicinity of Ga islands.

Finally, to demonstrate the capability of the correlative RISE, the maps of the Raman peak (G, 2D) shifts measured over the sample area are shown in Fig. 7. One can see the blue shifts of the Raman peaks across Ga islands and in their vicinity. The variations in Raman shifts for bare graphene can be for instance explained by an effect of doping [20] or by strain [21]. In [22] a Raman spectrum taken at the center of the bilayer graphene bubble is strongly blue-shifted compared to the Raman spectrum measured on the substrate. One can imaging Ga islands introduce the strain into graphene, similarly to these bubbles, and thus induce blue Raman shifts. This hypothesis is supported by our observations revealing that in the vicinity of Ga islands graphene partially detaches from the substrate due to an interplay of surface tension components at the Ga island perimeter (see Fig, S3). However, we do not want to speculate on these reasons in this paper and leave the exact explanation to further publications.

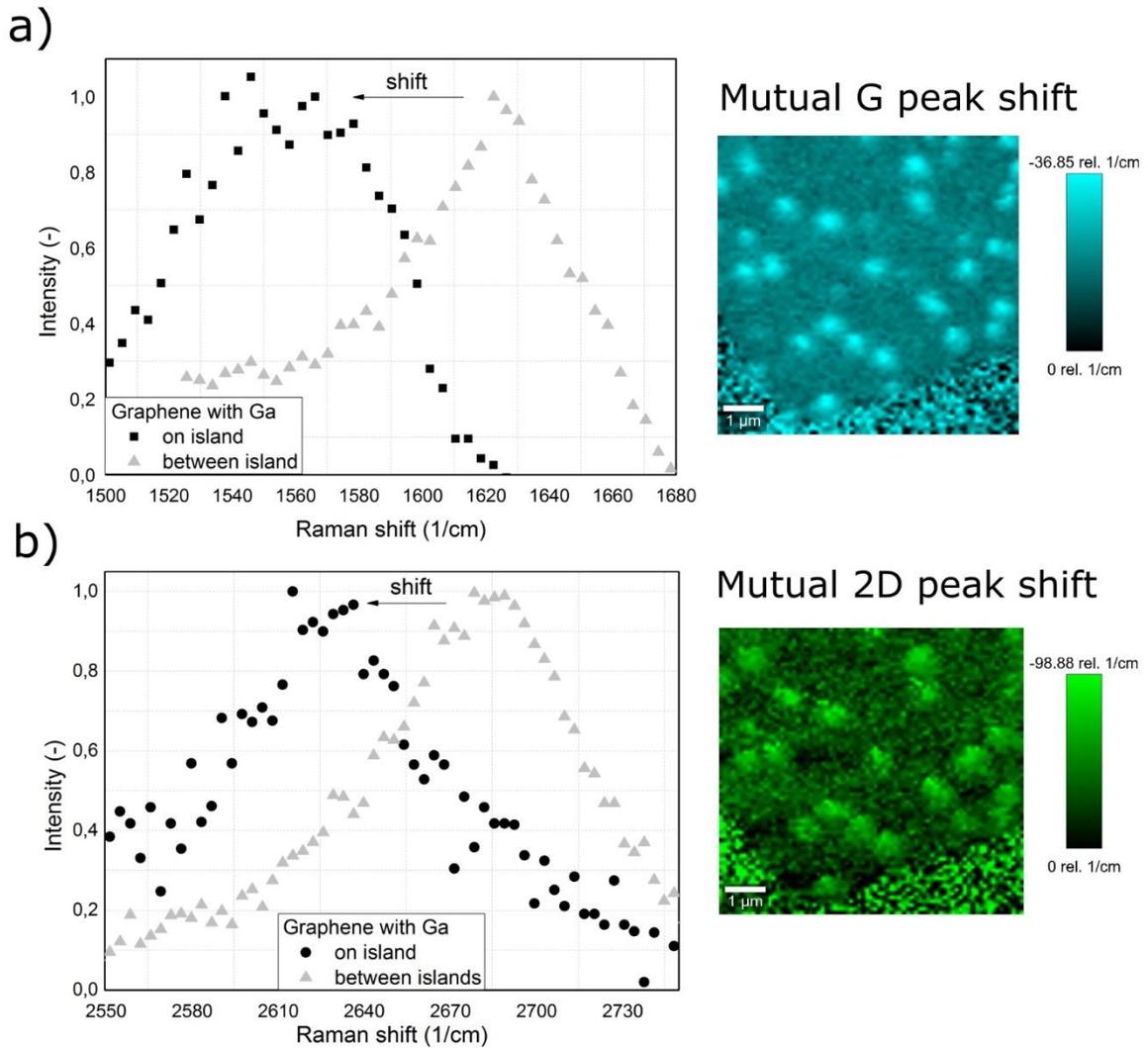

**Fig.7.** Raman shifts of the G and 2D peaks measured across the sample area.

**CONCLUSION**

Ga islands were deposited at different substrate temperatures on CVD graphene transferred on a Si(100) substrate with a 280 nm-thick $SiO_2$ layer. Depending on these temperatures, Ga islands of distinct diameters (40 – 300 nm), shape (spheres, disks), and spacing (up to 1 μm) were grown. Such a system exhibited graphene Raman peaks enhanced by individual Ga islands. Utilizing the correlative Raman imaging and SEM (RISE), it was found that this enhancement was island-size dependent and showed a maximum for medium-sized Ga islands. To prove the plasmonic nature of this effect, the enhancement of electric fields in the vicinity of Ga islands caused by localized surface plasmon resonances in these islands was simulated. A reasonable agreement between these simulations and the experimental intensities of Raman peaks confirms the plasmonic origin of the observed effect known as the Surface Enhanced Raman Spectroscopy.

**METHODS**

Single domain graphene flakes were grown by a standard low pressure CVD method [23] on a Cu substrate acting as a catalyst. In the first step of graphene growth, the reactor chamber with a copper substrate was pumped down to a base pressure of $10^{-3}$ Pa. Subsequently, it was filled with a hydrogen at a flow of 4 sccm and a pressure of 10 Pa, and the sample was annealed at $T = 1000$ °C. Then a $H_2/CH_4$ mixture was introduced into the furnace for 30 minutes at a flow of 40 sccm and a total pressure of 70 Pa to initiate graphene growth.

The graphene flakes were then transferred on a non-conductive $SiO_2$ (280nm) layer covering a *p*-doped Si(100) substrate (resistivity of $1.5 \times 10^{-3}$ Ωcm, provided by ON Semiconductor) using a PMMA-assisted wet transfer method [11]. After loading the sample into an UHV chamber with a base pressure of $p = 3 \times 10^{-8}$ Pa, the sample was annealed at $T = 300$°C for 24 hours to clean the graphene surface prior Ga deposition. The temperature was measured with an optical pyrometer (emissivity = 0.7). The annealing removes atmospheric adsorbates and PMMA residues which results in a decrease of graphene roughness. Ga atoms of thermal energy were evaporated by an effusion cell (Omicron EMF 3) from a PBN boat inside a Mo crucible. The Ga atom flux was kept at a constant value of $5.2 \times 10^{16}$ atoms per hour (0.2 ML/min.). After the deposition, the samples were characterized in situ by XPS and ex situ by AFM, and then by the correlative microscopy combining scanning electron microscopy (SEM) and Raman spectroscopy. The correlative Raman imaging and scanning electron microscopy (RISE) was carried out in a complex micro-spectroscopic setup (Tescan – RISE) equipped with a SEM (Lyra, Tescan), μRaman spectrometer providing a spatial resolution of 360 nm (Witec), and a piezoelectric stage (PI) enabling to position a sample in nanometer steps. The integration time for taking spectra at each point of the maps was 0.1 s, and the spot diameter and power of a 532 nm laser was 1 μm and 7mW, respectively. The objective (100 X) with a numerical aperture and working distance of 0.75 and 4 mm, respectively was used.

ACKNOWLEDGEMENT

We acknowledge the support by the Czech Science Foundation (Grant No.*20-28573S* ), European Commission (H2020-Twininning project No. 810626 – SINNCE, M-ERA NET HYSUCAP/TACR-TH71020004),*BUT* – specific research No.*FSI-S-20-648*5, and Ministry of Education, Youth and Sports of the Czech Republic (CzechNanoLab Research Infrastructure – LM2018110).

# Supplementary Information

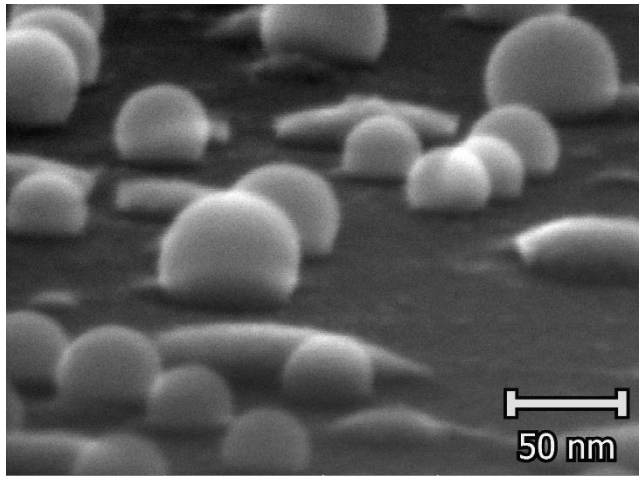

Fig. S1 SEM image showing two types of Ga nanostructures on graphene (tilted 80°).

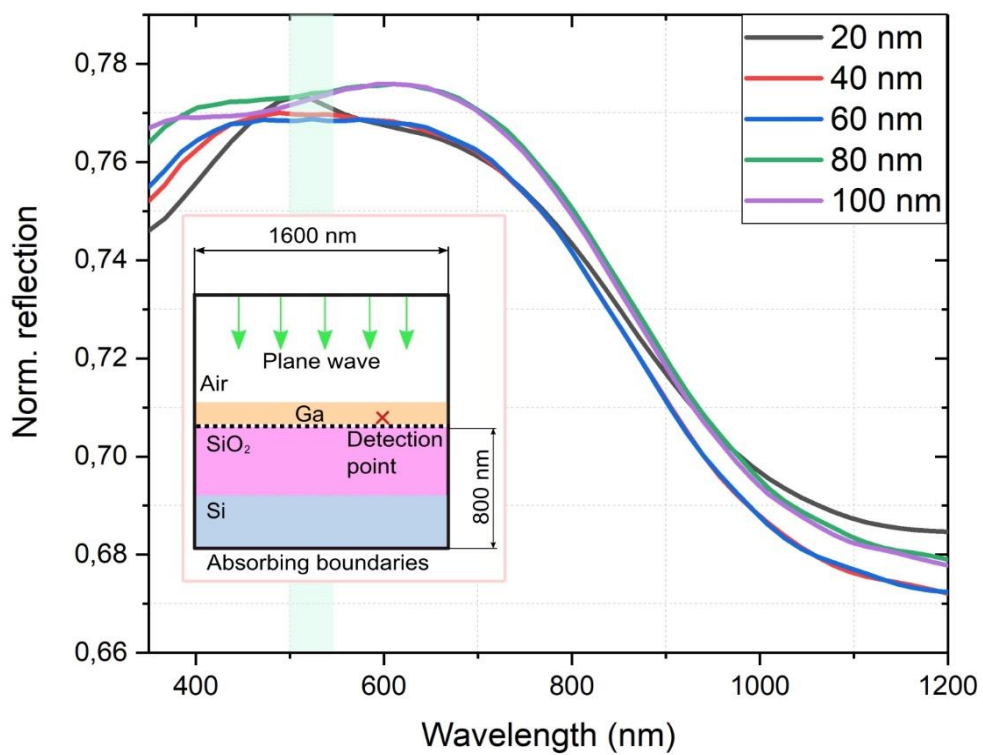

Fig. S2 Simulation of normalized reflection from a Ga/SiO$_2$/Si layered model (see the inset) as a function of the incident light wavelength for various thicknesses of the top Ga layer. There is no clear dependence on this thickness for any wavelength, including the laser one ($\lambda$ = 532 nm, see the dashed vertical line).

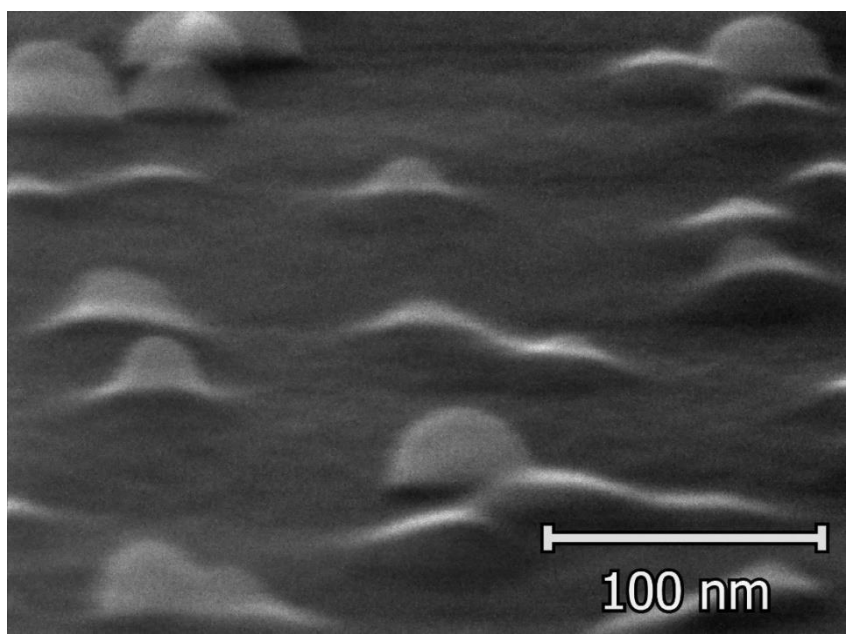

Fig. S3 SEM image indicating a detachment of graphene from the substrate in the vicinity of Ga islands.